# TITLE: ASSESSMENT OF SARS-COV-2 AIRBORNE INFECTION TRANSMISSION RISK IN PUBLIC BUSES

AUTHORS: M. Bertone[1], A. Mikszewski[2,3], L. Stabile[1,*], G. Riccio[4], G. Cortellessa[1], F.R. d'Ambrosio[5], V. Papa[6], L. Morawska[3], G. Buonanno[1,3]

[1] Department of Civil and Mechanical Engineering, University of Cassino and Southern Lazio, Cassino, FR, Italy

[2] CIUS Building Performance Lab, The City University of New York, New York, NY, USA

[3] International Laboratory for Air Quality and Health, Queensland University of Technology, Brisbane, Queensland, Australia

[4] Department of Industrial Engineering, University of Naples "Federico II", Italy

[5] Department of Industrial Engineering, University of Salerno, Italy

[6] Department of Motor Sciences and Wellness, University of Naples "Parthenope", Italy

*Corresponding author
Luca Stabile
Department of Civil and Mechanical Engineering
University of Cassino and Southern Lazio, Cassino
l.stabile@unicas.it

Abstract

Public transport environments are thought to play a key role in the spread of SARS-CoV-2 worldwide. Indeed, high crowding indexes (i.e. high numbers of people relative to the bus size), inadequate clean air supply, and frequent extended exposure durations make transport environments potential hotspots for transmission of respiratory infections. During the COVID-19 pandemic, generic mitigation measures (e.g. physical distancing) have been applied without also considering the airborne transmission route. This is due to the lack of quantified data about airborne contagion risk in transport environments.

In this study, we apply a novel combination of close proximity and room-scale risk assessment approaches for people sharing public transport environments to predict their contagion risk due to SARS-CoV-2 respiratory infection. In particular, the individual infection risk of susceptible subjects and the transmissibility of SARS-CoV-2 (expressed through the reproduction number) are evaluated for two types of buses, differing in terms of exposure time and crowding index: urban and long-distance buses. Infection risk and reproduction number are calculated for different scenarios as a function of the ventilation rates (both measured and estimated according to standards), crowding indexes, and travel times. The results show that for urban buses, the close proximity contribution significantly affects the maximum occupancy to maintain a reproductive number of < 1. In particular, full occupancy of the bus would be permitted only for an infected subject breathing, whereas for an infected subject speaking, masking would be required. For long-distance buses, full occupancy of the bus can be maintained only if specific mitigation solutions are simultaneously applied. For example, for an infected person speaking for 1 h, appropriate filtration of the recirculated air and simultaneous use of FFP2 masks would permit full occupancy of the bus for a period of almost 8 h. Otherwise, a high percentage of immunized persons (> 80%) would be needed.

1. Introduction

The COVID-19 pandemic, caused by the SARS-CoV-2 virus, has disrupted modern society and presented a significant challenge to indoor environments, where virus transmission mainly occurs (Blocken *et al.*, 2020; Chang *et al.*, 2021; Miller *et al.*, 2020; Morawska *et al.*, 2020; Wang *et al.*, 2020). Among the different pathways of infection transmission, the World Health Organization (WHO) and the US Centers for Disease Control and Prevention (CDC) (WHO, 30 April 2021; US CDC, 7 May 2021) have eventually recognized the airborne transmission of inhalable airborne respiratory particles (i.e. particles below 100 µm in diameter capable of remaining suspended in the air) as the dominant mode of respiratory infection in indoor environments with respect to sprayborne particles (larger particles quickly settling due to their inertia), and fomites (i.e. contaminated surfaces) (Marr and Tang, 2021; Miller *et al.*, 2020; Morawska and Milton, 2020; Kriegel *et al.*, 2020). Thus, indoor environments with a high crowding index (number of people relative to the room size) and inadequate clean (pathogen-free) air supply represent sites where the highest risk of microbial infection occurs (Miller *et al.*, 2020; Buonanno, Morawska and Stabile, 2020; Correia *et al.*, 2020; Li *et al.*, 2007). During the COVID-19 pandemic, the measures implemented by governments have not been targeted to reduce the transmission of the virus for all three mechanisms of transmission (airborne respiratory particles, sprayborne respiratory particles, and fomites). Indeed, the primary mitigation measures adopted have been physical distancing and hand hygiene, which address sprayborne respiratory particles and contaminated surfaces, but have limited effectiveness on transmission through airborne respiratory particles (Chen et al., 2020). This is due to the lack of data relating to quantified airborne contagion risk in indoor environments. Indeed, the first experimental evidence of SARS-CoV-2 RNA concentration in indoor air in the presence of an infected person (Lednicky *et al.*, 2020; Liu *et al.*, 2020; Nissen *et al.*, 2020; Stern *et al.*, 2021) as well as traces of SARS-CoV-2 RNA on air conditioning filters and ambient air in buses (Moreno et al., 2021) were only recently reported in the scientific literature. Thus, given the millions of commuters using public transport every day in the world, there is an obvious need for further information on the risk of airborne contagion so that effective prevention measures can be implemented. In particular, airborne transmission of SARS-CoV-2 on buses has emerged as a concern in light of multiple outbreaks since the onset of the pandemic (Luo *et al.*, 2020; Shen *et al.*, 2020). Buses have historically been associated with transmission of *Mycobacterium tuberculosis* (Mohr *et al.*, 2012) and measles virus (Perkins, Bahlke and Silverman, 1947; Helfand *et al.*, 1998), with one case report for variola (smallpox) virus (Suleimanov and Mandokhel, 1972). Therefore, mitigation of airborne transmission of respiratory pathogens on buses represents an important topic even beyond the COVID-19 pandemic.

The significance of airborne transmission has highlighted the need to have appropriate pathogen-free air supply rates (i.e. air exchange rates) to reduce the spread of SARS-CoV-2 (Morawska *et al.*, 2021; Buonanno, Morawska and Stabile, 2020; Stabile *et al.*, 2021). Despite numerous studies quantifying air exchange rates in indoor microenvironments, no existing ventilation standard so far developed by national authorities or international professional societies (e.g. ASHRAE 62.1 [2019]) takes into consideration the requirements for infection control in non-healthcare settings (Morawska *et al.*, 2021). In addition, there are no specific technical regulations or standards for buses focused on ventilation and air exchange rate. Regulation n°107 of the UNECE (United Nations Economic Commission for Europe) on the uniform provisions concerning the approval of vehicles, including buses, does not consider the ventilation systems and defines the maximum bus capacity considering only the available internal surface and the maximum permissible load. Nevertheless, some countries have a standard for buses; for example the German standard VDV 236 (2015) requires 15 m³ h$^{-1}$ person$^{-1}$ of clean air, and the Chinese standard JT/T 888 (2014) requires 20 m³ h$^{-1}$ person$^{-1}$ of clean air.

The purpose of this study was to quantify the risk of airborne transmission in buses and identify mitigation strategies to reduce the transmission potential of SARS-CoV-2 infection for safe transportation of passengers and to control the spread of the pandemic. To this end, we performed risk assessment simulations considering airborne transmission both in close proximity to an infected passenger (i.e. within 1.5 m) and at more distant locations in the bus breathing shared air (referred to as "room-scale"). For the close proximity component, we applied a computational fluid dynamics (CFD) approach, whereas for the room-scale component we applied a simplified zero-dimensional model based on a virus mass balance that allows prospective analyses. Simulations were performed considering various exposure scenarios in the bus environment taking into account the characteristics of the emitting subject, microenvironment, ventilation, and exposed subjects also including the effect of mitigation strategies. As the Delta variant (B.1.617.2 SARS-CoV-2) is now dominant across much of the world, and is recognized as more infectious than previous variants, the risk assessment proposed here focuses on this variant.

## 2. Materials and Methods

To quantify the risk of airborne transmission of viruses in buses and to identify mitigation strategies to reduce the transmission potential of SARS-CoV-2 infection for safe transportation of commuters, the approaches proposed by Cortellessa *et al.* (2021) and Buonanno, Morawska and Stabile (2020) are used. In particular, we evaluate airborne transmission resulting from inhaling virus-laden airborne particles at two different spatial scales: i) in *close proximity*, i.e. within approximately 1.5 m of an emitting subject; and ii) at *room-scale*, i.e. sharing the same indoor environment of the infected subject and then inhaling particles that remain suspended in air. The dichotomy of close proximity versus room-scale airborne transmission has also been referred to as short-range versus long-range transmission (Chen et al., 2020). Individual risk of infection (i.e. the ratio between the number of new infections and the number of exposed susceptible individuals) and reproductive number (i.e. the expected number of new infections arising from a single infectious individual) for both close proximity and room-scale are evaluated adopting an exposure-to-risk approach developed and presented in our previous papers (Buonanno *et al.*, 2020; Buonanno, Morawska and Stabile, 2020; Buonanno, Stabile and Morawska, 2020; Stabile *et al.*, 2021). This approach is summarized and customized for close proximity and room-scale assessments. Different scenarios are studied with ventilation rates estimated according to regulatory standards and measured through an *ad-hoc* experimental campaign and crowding indexes required by regulatory authorities.

### 2.1. Evaluation of the individual risk of infection and reproductive number for close proximity transmission

The close proximity approach consists of a Eulerian-Lagrangian based model for the analysis of respiratory particle dispersion in close proximity represented by a breathing/speaking infected subject (emitter) and a susceptible subject (receiver) in the case of face-to-face orientation and stagnant air conditions. A CFD technique is adopted for the three-dimensional numerical description of velocity, pressure, and temperature fields, along with the motion and interaction of the respiratory particles with the fluid. The fully opensource finite volume based OpenFOAM software is employed as a fully open and flexible tool with complete control of the variables chosen for particle dispersion assessment. The adopted Lagrangian particle tracking (LPT) approach is based on a dispersed dilute two-phase flow and allows the respiratory particle motion inside the air flow to be determined. In particular, the spacing between respiratory particles in the exhaled air plume is sufficiently large and the volume fraction of the respiratory particles is sufficiently low (< $10^{-3}$) to justify the use of a Eulerian-Lagrangian approach, in which the continuum equations are solved for the air flow (continuous phase) and Newton's equation of motion is solved for each respiratory particle. The continuity equations are widely described in the available scientific literature (Arpino *et al.*, 2014; Massarotti *et al.*, 2006; Scungio *et al.*, 2013) while the respiratory particle motion equations, solved for an unsteady incompressible Newtonian fluid and considering the drag and gravity forces acting on the particle, are described in Cortellessa *et al.* (2021) and not reported here for brevity. Further details are reported in the Supplementary Material where the particle emission rates as a function of the particle size for breathing activity are also summarized (Table S1).

The close proximity airborne transmission risk is evaluated using the volumetric dose of airborne particles pre-evaporation ($V_{d\text{-airborne-pre}}$) inhaled by a susceptible person during face-to-face interaction with an infected person. We use the pre-evaporation volume because the dose of RNA copies inhaled relates to the original volume rather than the evaporated volume, as particles retain their RNA load while losing water during the instantaneous evaporation occurring upon expiration. This dose of inhaled RNA copies can be approximated as the product of $V_{d\text{-airborne-pre}}$, the viral load ($c_v$) of the infected person, and the duration of face-to-face interaction in minutes. Values for $V_{d\text{-airborne-pre}}$ when the infected person is speaking are taken from Cortellessa *et al.* (2021) for separation distances of up to 1.75 m, whereas original estimates for $V_{d\text{-airborne-pre}}$ when the infected person is breathing only for separation distances of up to 0.5 m are reported in the Supplementary Material (Table S2). CFD models for both speaking and breathing activities do not consider mask use because masks completely alter the close proximity particle flow regime, thus reducing the close proximity risk. In our calculation we consider that the close proximity risk is negligible in a scenario where the infected person wears a mask.

To model $c_v$, we used the preliminary data posted by von Wintersdorff *et al.* (2021) that confirms that higher $c_v$ values are associated with Delta variant infections; in particular, we fit a lognormal $c_v$ distribution to the approximate interquartile range of sequence-confirmed Delta variant infection data only (n = 87), yielding a mean and standard deviation of 7.1 and 0.70 $\log_{10}$ RNA copies mL$^{-1}$, respectively (von Wintersdorff et al., 2021).

To calculate the probability of infection ($P_I$) from close proximity airborne transmission, we used a common exponential dose-response model as follows:

$$P_I = 1 - e^{-\frac{C_v V_{d-airborne-pre}}{\text{HID}_{63}}} \quad (\%) \quad (1)$$

where $\text{HID}_{63}$ represents the human infectious dose for 63% of susceptible subjects. For the Delta variant, a $\text{HID}_{63}$ value of 700 RNA copies was adopted based on the thermodynamic equilibrium dose-response model of Gale (2020). We point out that the term ($c_v V_{d\text{-airborne-pre}}/\text{HID}_{63}$) represents the term hereafter referred to as "dose of quanta" ($D_q$).

The close proximity individual risk of infection ($IR_{cp}$) of the exposed person was then calculated by integrating, for all the possible $c_v$ values, the product between the conditional probability of the infection for each $c_v$ ($P_I(c_v)$) and the probability of occurrence of each $c_v$ value ($P_{cv}$):

$$IR_{cp} = \int_{c_v} (P_I(c_v) * P_{cv}) dc_v \qquad (\%) \qquad (2)$$

For the purposes of our modeling analysis, we assume an infected person on a bus has close proximity interaction (speaking or breathing) with only one susceptible person; in other words, just one susceptible person is within 1.5 m and in a face-to-face orientation for the whole exposure time (i.e. travel time). As a result, the close proximity reproduction number ($R_{cp}$) (i.e. the number of secondary cases amongst the susceptibles in close proximity to an infected subject) is equivalent to the close proximity individual risk of infection ($IR_{cp}$).

## 2.2. Evaluation of the individual risk of infection and reproductive number for room-scale transmission

The room-scale approach is based on a box model in which a virus mass balance equation is applied, estimating the emission of an infected subject and predicting exposure concentrations and infection risks for prospective scenarios. The approach is based on the following hypotheses: the emitted particles are instantaneously and evenly distributed in the environment, and the latent period of the disease is longer than the time of the model (Gammaitoni and Nucci, 1997). Infected people breathing and/or speaking and susceptible people standing are considered.

For room-scale airborne transmission assessment, the predictive estimation approach developed by Buonanno *et al.* (2020), and already applied in Buonanno, Morawska and Stabile (2020), Buonanno, Stabile and Morawska (2020), Moreno *et al.* (2021) and Stabile *et al.* (2021), was adopted. The approach requires six steps: (i) evaluation of the quanta emission rate; (ii) estimation of the exposure to quanta concentration in the environment; (iii) evaluation of the dose of quanta received by exposed subjects; (iv) estimation of the probability of infection based on a dose-response model; (v) evaluation of the individual risk of the exposed person; and (vi) evaluation of the room-scale reproduction number ($R_{rs}$) based on crowding. The abovementioned quantum is defined as an inhaled dose of RNA copies of SARS-CoV-2 that can cause infection in 63% of susceptible people in an indoor environment, whereas the quanta emission rate is the number of quanta released into the air per unit of time as a function of the expiratory activities of an infected subject, respiratory parameters, and activity levels. The approach estimates the quanta emission rate of an infectious subject based on the viral load in the respiratory fluid and the concentration of particles expired during different activities; moreover, it considers the metabolic rate and respiratory activity of the emitting subject and the activity of the exposed subject.

This approach represents an important step forward, as previously the viral load emitted was difficult to estimate; in fact, a backward calculation was used to estimate the emission of an infected subject based on retrospective assessments of outbreaks only at the end of an epidemic (Myatt *et al.*, 2008; Rudnick and Milton, 2003; Sze To and Chao, 2010; Wagner, Coburn and Blower, 2009).

The quanta emission rate ($ER_q$, quanta h$^{-1}$) is evaluated as:

$$ER_q = c_v \cdot c_i \cdot IR \cdot V_d \qquad (\text{quanta}^{-1}) \qquad (3)$$

where $c_v$ (RNA copies mL$^{-1}$) is the viral load in the saliva (as defined in Section 2.1), $c_i$ (quanta RNA copies) is a conversion factor defined as the ratio between one infectious quantum and the infectious dose expressed in viral RNA copies (assumed as in Section 2.1), IR is the inhalation rate (m$^3$ h$^{-1}$) of the exposed subject, which is a function of the subject's activity level and age, and $V_d$ is the droplet volume concentration expelled by the infectious person (mL m$^{-3}$). IR and $V_d$ data are reported in Buonanno, Morawska and Stabile (2020). The resulting quanta emission rate distribution (ER$_q$, quanta h$^{-1}$), expressed as $\log_{10}$ (average ± standard deviation), are 1.19 ± 0.68 and 1.84 ± 0.68 for oral breathing and speaking, respectively.

The indoor quanta concentration in the environment $n(t,ER_q)$ is evaluated for each possible ER$_q$ value using the equation:

$$n(t, ER_q) = n_0 \cdot e^{-IVRR \cdot t} + \frac{ER_q \cdot I}{IVRR \cdot V} \cdot (1 - e^{-IVRR \cdot t}) \qquad \text{(quanta m}^{-3}\text{)} \qquad (4)$$

where $n_0$ (quanta m$^{-3}$) is the initial quanta concentration in the bus (assumed to be zero), IVRR (h$^{-1}$) represents the infectious virus removal rate and is the sum of three contributions (Yang and Marr, 2011): (i) AER (h$^{-1}$), the air exchange rate; (ii) k (h$^{-1}$), the particle deposition rate on surfaces (equal to 0.24 h$^{-1}$, (Chatoutsidou and Lazaridis, 2019)); (iii) λ (h$^{-1}$), the viral inactivation rate (equal to 0.63 h$^{-1}$, (van Doremalen *et al.*, 2020)); *I* is the number of infectious subjects, and V is the volume of the buses considered.

The dose of quanta received by an exposed subject ($D_q$) to a certain quanta concentration, $n(t,ER_q)$, for a certain exposure time, *t*, can be evaluated by integrating the quanta concentration over time as:

$$D_q(ER_q) = IR \cdot \int_0^t n(t)dt \qquad \text{(quanta)} \qquad (5)$$

The probability of infection ($P_I$, %) of exposed persons is evaluated based on the same exponential dose-response model considered for close proximity:

$$P_I = 1 - e^{-D_q} \qquad \text{(\%)} \qquad (6)$$

Once again, we point out that close proximity and room-scale approaches are based on the same exposure-to-risk evaluation; indeed, eq. 6 is practically the same as eq. 1.

The room-scale individual infection risk of an exposed person (IR$_{rs}$) is calculated by integrating, for all the possible ER$_q$ values, the product between the conditional probability of the infection for each ER$_q$ ($P_I(ER_q)$) and the probability of occurrence of each ER$_q$ value ($P_{ERq}$):

$$IR_{rs} = \int_{ER_q} R(ER_q) dER_q = \int_{ER_q} (P_I(ER_q) \cdot P_{ER_q}) dER_q \qquad \text{(\%)} \qquad (7)$$

The room-scale reproduction number (R$_{rs}$) represents the expected number of secondary cases arising from the exposure and is simply calculated as the product of IR$_{rs}$ and the number of susceptible passengers on the bus. When considering both close proximity and room-scale airborne transmission, the total number of expected secondary cases arising from the bus trip (R$_{event}$) can be approximated as the sum of R$_{cp}$ and R$_{rs}$.

With a view towards minimizing the spread of infection, such that the bus exposure results in fewer than one secondary transmission on average, the number of susceptible passengers should be monitored to maintain a condition where $R_{event}$ < 1. To this end, the maximum number of susceptibles that can stay simultaneously in the confined space under investigation for an acceptable $R_{event}$ < 1 (hereafter referred to as maximum room occupancy, MRO), considering a single close proximity interaction and room-scale airborne transmission, is:

$$MRO = \frac{1-IR_{cp}}{IR_{rs}} \quad \text{(susceptibles)} \quad (8)$$

### 2.3. Scenarios

The individual infection risk of susceptible subjects and the overall transmission potential (expressed as reproduction number) are evaluated for two types of buses, differing in terms of exposure time and crowding index, as follows: i) urban buses (class I) characterized by a short exposure time and high crowding index and ii) long-distance buses (class II and class III) characterized by long exposure time and a low crowding index. The bus classes are defined by the Regulation n°107 of UNECE on uniform provisions concerning the approval of vehicles, including buses, as shown in Table 1. In class II buses, standing passengers are allowed, although they are unlikely to carry standing people over long distances, making class II and III buses practically identical. Thus, only class III buses were considered in the simulation of long-distance buses.

Table 1 - Classes of buses according to Regulation n°107 of the UNECE (ECE-R107, 2015).

| | |
|---|---|
| Class I buses | Vehicles constructed with areas for standing passengers, to allow frequent passenger movement |
| Class II buses | Vehicles constructed principally for the carriage of seated passengers and designed to allow the carriage of standing passengers in the gangway and/or in an area that does not exceed the space provided for two double seats |
| Class III buses | Vehicles constructed exclusively for the carriage of seated passengers |

Infection risk and reproduction numbers are calculated for different scenarios, with ventilation rates estimated according to regulatory standards as well as being measured through an *ad-hoc* experimental campaign, and crowding indexes required by regulatory authorities. For long-distance buses with long exposure times (≥ 120 min), the room-scale airborne transmission risk will dominate and act to reduce the MRO and consequently increase the effective distances between passengers. For reference, the close proximity risks are negligible beyond 1.75 m in the case of speaking and 0.5 m in the case of breathing (see Supplemental Material). Conversely, for the short exposure time and high crowding of class I buses, the close proximity risk may affect the MRO. As such, the close proximity risk was only specifically considered for urban bus scenarios and, on the basis of the assumption mentioned above, it is negligible for scenarios where commuters wear masks (and thus it has not been considered).

#### 2.3.1. Simulated scenarios

The individual infection risk of susceptible subjects and the reproduction number are evaluated for two types of buses: urban buses (class I) and long-distance buses (class III). The individual bus capacity depends on the size of the vehicle, the seating configuration, and the regulation regarding standees (Table 2). In the present work, we consider a widely-used conventional bus (Victor and Ponnuswamy, 2012), approximately 12 meters long and with a maximum passenger capacity depending on the class in which it is used. The bus dimensions considered in this work are 12 m × 2.55 m × 2.3 m (L × W × H) equal to a volume of 70 m$^3$. For accurate calculation of the actual internal volume, the volume occupied by the seats is removed as well as the volume occupied by passengers. The crowding index, suggested by Regulation n°107 (ECE-R107, 2015) on the uniform provisions concerning the approval of vehicles, including buses, amounts to 93 and 51 occupants (seated + standees, excluding the driver) for class I and class III buses, respectively. Considering a density of humans of 1010 kg m$^{-3}$ (Deziel, 2021), the mass of a passenger of 68 kg (ECE-R107, 2015), and the crowding index expected by Regulation n°107 (ECE-R107, 2015), the volume occupied by passengers is 6 m$^3$ and 3 m$^3$ for class I and class III, respectively. Moreover, the volume occupied by the seats expected by Regulation n°107 (ECE-R107, 2015) is equal to 1.4 m$^3$ and 2 m$^3$, for class I and class III, respectively.

The heating, ventilation and air conditioning (HVAC) system plays a key role in the airborne transmission of respiratory infections within a bus because it adds clean, pathogen-free air to the indoor environment as a fraction of its flow rate. The data on buses and HVAC systems have been found by consulting technical data sheets provided by manufacturers. The considered HVAC system can provide a maximum flow rate of 4400 m$^3$ h$^{-1}$ including both recirculated air and outdoor fresh air. Detailed legislation on the minimum outdoor fresh air to be supplied in buses is missing, thus the reference value for urban and suburban rolling stock, suggested by EN 1432-1 (2006), is typically adopted: it is equal to 15 m$^3$ h$^{-1}$ person$^{-1}$ (i.e.~4.17 L s$^{-1}$ person$^{-1}$) or 22 h$^{-1}$ and 12 h$^{-1}$ for class I and III, respectively (Table 2). The recirculated air should be treated using filters able to capture and remove particles with different efficiencies depending on the filter used. If this does not occur, the recirculated air will not enhance the air exchange rate guaranteed by the outdoor air supply. In the simulations performed here, we considered three different efficiencies for filtering the recirculated air: no filtration, filtration through a G3 filter, and filtration through a M6 filter (ISO 16890-1, 2016). All the simulations for urban buses were carried out with no filtration of the recirculated air. Considering the distribution of the particles emitted by an infected subject during speaking (distribution post-evaporation fitted by seven size ranges as reported in Cortellessa *et al.*, 2021), and the efficiency declared by regulation ISO 16890-1 (2016), the filters guarantee a weighted average removal efficiency of 4% for G3 and 40% for M6 filters, respectively.

The actual air exchange rate in buses is also affected by the opening of windows and doors, which increases the air exchange rate. This typically occurs in urban buses, which are characterized by frequent stops, when doors must be opened, and high crowding indexes leading to windows being kept open. For this reason, the actual ventilation rate for urban buses (class I) were measured through an *ad-hoc* experimental campaign as described in section 2.3.2.

Table 2 – Characteristics of the buses in terms of maximum occupancy, volume, crowding index, and ventilation rate.

| Bus class | Maximum occupancy suggested by the (ECE-R107, 2015) regulation | | | Volume (m$^3$) | Crowding index (person m$^{-3}$) | Air exchange rate due to outdoor fresh air (EN 1432-1) (h$^{-1}$) |
|---|---|---|---|---|---|---|
| | Seats | Standees | Tot | | | |
| I | 36 | 57 | 93 | 63 | 1.5 | 22 |
| III | 51 | - | 51 | 65 | 0.8 | 12 |

The simulations are performed considering one infected passenger (I = 1) in a fully susceptible population. The exposed susceptibles were considered to be performing activities in sitting and standing positions and inhaling at IR = 0.54 m$^3$h$^{-1}$ (Adams, 1993 and ICRP, 1994). Travel times on buses vary widely between urban and long-distance buses depending on the number of stops, travel distance, and traffic patterns. The travel time considered is: (i) 24 min for urban buses (class I), which is the average time spent by a commuter on an urban bus in Italy (ISTAT, 2019); and (ii) travel time up to 8 h for long-distance buses (class III). The urban bus scenarios were simulated considering that all the 93 commuters stayed simultaneously for 24 min in the bus. This is a rare case, but it represents the worst situation and should be considered as a conservative approach.

Our modeling scenarios also consider the use of face masks by bus passengers, because universal masking has been a primary public health strategy during the COVID-19 pandemic. In the scenarios with this mitigation solution, all the commuters wear a mask, both infected and susceptible, and we consider both surgical and FFP2 masks. For the surgical masks, we assume a 40% reduction in inhaled particles (Eikenberry *et al.*, 2020), seen as the product of the reduction of the emission of the infected subject and the inhalation of the susceptibles. For the FFP2 masks, the overall considered reduction effect was assumed to be 80% (Poydenot *et al.*, 2021).

Emission and exposure assumptions for the scenarios in the prospective assessment for urban buses (class I) and long-distance buses (class III) are summarized in Table 3 and Table 4, respectively. For all the scenarios adopted, the commuter was considered to be an emitting subject.

Table 3 - Scenarios simulated for urban buses (class I): emission duration and respiratory activity. Descriptions of the scenarios and the activity mitigation solutions are reported. All the simulations were carried out with the actual air exchange rates and with no filtration of the recirculated air.

| Scenarios | | Emission duration (min) and respiratory activity | Description |
|---|---|---|---|
| Base scenario | C-0-UB | 24 min, oral breathing | Infected commuter standing for the whole trip oral breathing. No filtration of the recirculated air. |
| Speaking effect & windows closed | C-24-UB | 24 min, speaking | Infected commuter standing for the whole trip speaking and with windows closed. No filtration of the recirculated air. |
| Speaking effect & windows opened | C-24-UB-WO | 24 min, speaking | Infected commuter standing for the whole trip speaking and with windows opened. No filtration of the recirculated air. |
| Surgical mask, speaking effect & windows closed | C-24-UB-SM | 24 min, speaking | Infected commuter standing for the whole trip speaking, all commuters wearing a surgical mask and with windows closed. No filtration of the recirculated air. |
| FFP2 mask, speaking effect & windows closed | C-24-UB-FFP2 | 24 min, speaking | Infected commuter standing for the whole trip speaking, all commuters wearing an FFP2 mask and with windows closed. No filtration of the recirculated air. |
| Surgical mask, windows open & speaking effect | C-24-UB-SM+WO | 24 min, speaking | Infected commuter standing for the whole trip speaking, all commuters wearing a surgical mask and with windows opened. No filtration of the recirculated air. |
| FFP2 mask, windows open & speaking effect | C-24-UB-FFP2+WO | 24 min, speaking | Infected commuter standing for the whole trip speaking, all commuters wearing an FFP2 mask and with windows opened. No filtration of the recirculated air. |

Table 4 - Scenarios simulated for long-distance buses (class III): emission duration and respiratory activity. Descriptions of the scenarios and the activity mitigation solutions are reported. All the simulations of the long-distance buses were carried out with the air exchange rates suggested by the standard.

| Scenarios | | Emission duration (min) and respiratory activity | Description |
|---|---|---|---|
| Base scenario | C-0-LDB | 480 min oral breathing | Infected commuter standing for the whole trip oral breathing. No filtration of the recirculated air. |
| Commuter's speaking effect | C-30-LDB | 30 min speaking & 450 min oral breathing | Infected commuter speaking for the first 30 min and oral breathing for the rest of the time. No filtration of the recirculated air. |

| | C-60-LDB | 60 min speaking & 420 min oral breathing | Infected commuter speaking for the first 60 min and oral breathing for the rest of the time. No filtration of the recirculated air. |
|---|---|---|---|
| Filtration G3 & speaking effect | C-60-LDB-G3 | 60 min speaking & 420 min oral breathing | Infected commuter speaking for the first 60 min and oral breathing for the rest of the time. The recirculated air is filtered with a G3 filter. |
| Surgical mask effect | C-0-LDB-SM | 480 min oral breathing | Infected commuter standing for the whole trip oral breathing. All commuters wear surgical masks. No filtration of the recirculated air. |
| Filtration M6 effect | C-0-LDB-M6 | 480 min oral breathing | Infected commuter standing for the whole trip oral breathing. The recirculated air is filtered with an M6 filter. |
| Filtration M6 & speaking effect | C-60-LDB-M6 | 60 min speaking & 420 min oral breathing | Infected commuter speaking for the first 60 min and oral breathing for the rest of the time. The recirculated air is filtered with an M6 filter. |
| Surgical mask & speaking effect | C-60-LDB-SM | 60 min speaking & 420 min oral breathing | Infected commuter speaking for the first 60 min and oral breathing for the rest of the time. All commuters wear surgical masks. No filtration of the recirculated air. |
| FFP2 & speaking effect | C-60-LDB-FFP2 | 60 min speaking & 420 min oral breathing | Infected commuter speaking for the first 60 min and oral breathing for the rest of the time. All commuters wear an FFP2 mask. No filtration of the recirculated air. |
| FFP2 & filtration M6 effect | C-60-LDB-FFP2+M6 | 60 min speaking & 420 min oral breathing | Infected commuter speaking for the first 60 min and oral breathing for the rest of the time. The recirculated air is filtered with an M6 filter and all commuters wear a surgical mask. |

### 2.3.2. Measurement of the air exchange rate for buses class I

The actual air exchange rate for urban buses (class I) was determined through in-field measurements using the decay method of a tracer gas (Van Buggenhout *et al.*, 2009; Cui *et al.*, 2015; ISO 12569, 2017). In brief, a dose of tracer gas is injected and mixed with the air inside the bus. When the injection is stopped, the concentration peak is reached and the tracer gas concentration begins to decrease and is recorded during a given period. The tracer gas decay method is based on the mass balance of the tracer gas which allows the air exchange rate (AER, $h^{-1}$) to be calculated through the exponential decay equation:

$$AER = \frac{ln\frac{(C_{peak}-C_{out})}{(C_{final}-C_{out})}}{\Delta t} \quad (h^{-1}) \quad (9)$$

where $C_{peak}$, $C_{final}$, and $C_{out}$ represent the initial peak, final, and outdoor tracer gas concentrations, respectively, and $\Delta t$ the time interval between $C_{peak}$ and $C_{final}$.

In the experimental campaign, carbon dioxide ($CO_2$) was used as a tracer gas. Measurements were conducted with six Onset HOBO MX1102 $CO_2$ data loggers (Range 0 to 5000 ppm $CO_2$, accuracy ± 50 ppm ± 5%) and a Testo 435 multifunctional logger with IAQ (Range 0 to 10000 ppm - Accuracy: 0 to 5000 ppm $CO_2$, ± 75 ppm ± 3%; 5000 to 10000 ppm $CO_2$, ± 150 ppm ± 5%). Before the measurement campaign, the sensors were calibrated by using pure nitrogen and an analytically calibrated gas mixture with a concentration of 4000 ppm $CO_2$ in nitrogen.

The measurements were carried out in two buses used for urban transport, namely IIA CityMood bus (class I) on an ordinary route and the measurements were carried out both with the windows closed and the windows open. The ventilation system of the buses was kept in operation for the entire duration of the tests. The tracer gas was fed from a cylinder, until a $CO_2$ concentration peak of about 5000 ppm was reached, then the tracer supply was interrupted, and the concentration decay was recorded. The $CO_2$ concentration was measured at several points in the bus because the variable vehicle speed and the repeated opening of doors did not allow the tracer gas concentration to achieve uniformity. The sensors were placed at two heights corresponding to the breathing zones for seated (1.10 m) and standing (1.70 m) passengers. Evaluation of the AER value was carried out based on the $CO_2$ concentration measured during the first phase of the decay when the $CO_2$ level was high and the occupants' contribution was negligible. The $CO_2$ concentration values measured at each point were processed according to the decay method to calculate the local value of the air exchange rate and, finally, the average value of AER was calculated. The measurements collected in the experimental campaign are shown in Table 5. The data clearly show that for the windows closed condition, the minimum AER due to outdoor fresh air suggested by the EN 1432-1 standard (22 $h^{-1}$, i.e. 4.17 L $s^{-1}$ $person^{-1}$) is guaranteed; indeed, the experimental AER value with windows closed is slightly larger than the value suggested by the standard, likely due to the frequent stops of the bus during which the doors were opened. However, when the windows were kept open during the whole trip, the actual AER was roughly three times the prescribed value, reaching 65.3 $h^{-1}$.

Table 5 - Actual air exchange rates (AER) measured for urban bus IIA CityMood bus (class I) during the experimental analyses. Data are reported as average ± standard deviation values.

| Experiment | AER ($h^{-1}$) | Ventilation rate per person (L $s^{-1}$ $person^{-1}$) |
|---|---|---|
| Windows open | 65.3 ± 4.6 | 12.4 ± 0.9 |
| Windows closed | 26.9 ± 3.9 | 5.1 ± 0.7 |

## 3. Results and discussions

### 3.1 Urban buses

Figure **1** presents the close proximity and room-scale individual risks after 24 min for the breathing (C-0-UB) and speaking (C-24-UB) scenarios. The room-scale risk in the case of windows open for enhanced ventilation (C-24-UB-WO) is also shown. For the speaking scenario, the close proximity risk exceeds the room-scale risk for separation distances below approximately 1.5 m; in particular, the close-proximity individual risk is extremely high (~75%) in the case of full occupancy of the bus (93 persons, which means an average separation distance of 0.32 m). Because we assumed that an infected person has close proximity interaction with only one susceptible person, the maximum $R_{cp}$ value (resulting from a separation distance of 0.32 m) is about 0.75. Conversely, for the breathing scenario, the close proximity risk is only higher than the room-scale one when the infected person is within 0.2 m of a susceptible person, and the close proximity risk is very low (~0.2%) at the designed distance of 0.32 m. As such, the close proximity risk for breathing has a minimal impact on the reproduction number for the scenarios evaluated here, and can be omitted from all $R_{event}$ calculations. Indeed, in the case of an infected subject breathing for the entire trip, the $IR_{rs}$ is 0.48%, and the $R_{rs} = R_{event} = 0.44$ (i.e. calculated as the product between 0.48% and the 92 susceptibles); therefore, the full occupancy suggested by the ECE-R107, 2015 regulation (93 commuters) is satisfied.

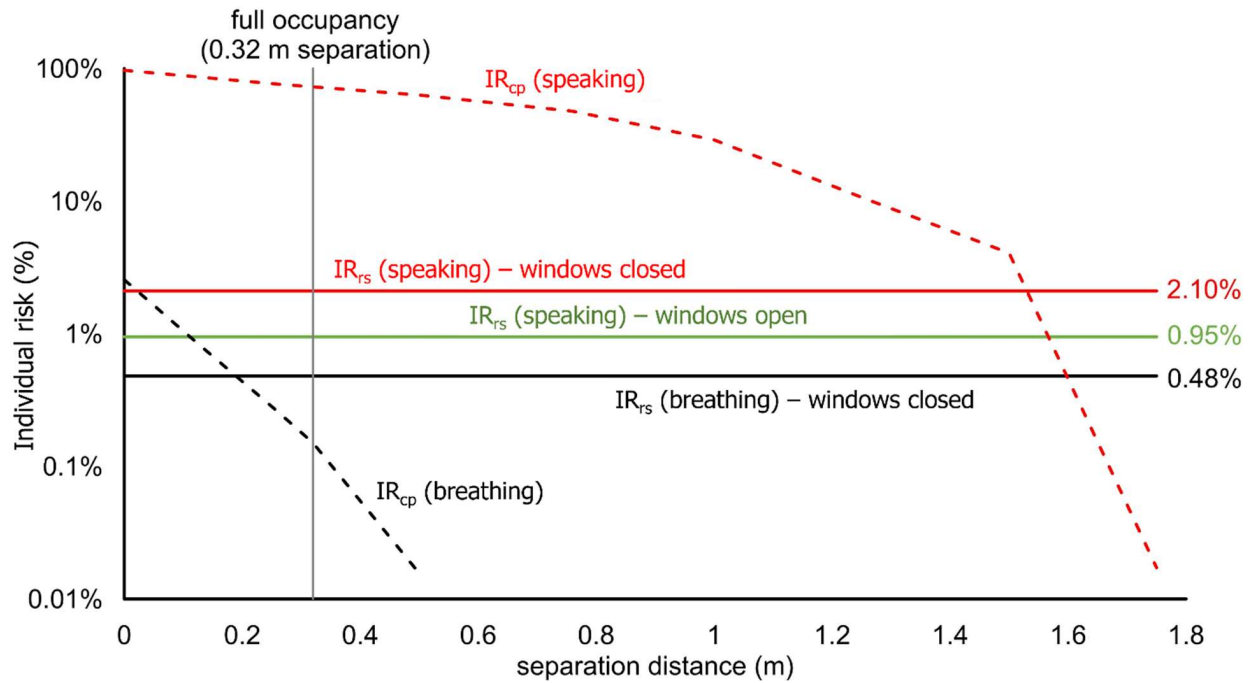

**Figure 1** - Close proximity risk ($IR_{cp}$) as a function of separation distance for the Delta variant for 24 min of breathing and speaking, as compared with room-scale airborne transmission risk ($IR_{rs}$) for the C-0-UB (breathing, windows closed), C-24-UB (speaking, windows closed), and C-24-UB-WO (speaking, windows open) scenarios.

In contrast with the breathing scenarios, for an infected subject speaking during the whole trip, the $R_{event}$, and then the MRO, is also significantly affected by the close proximity risk of infection due to its high $IR_{cp}$. The effect of close proximity to the total $R_{event}$ becomes even more predominant when the number of susceptibles decreases (i.e. when the $R_{rs}$ decreases). This is clearly shown in Figure 2, where the $R_{cp}$ was added to the $R_{rs}$ to calculate the total $R_{event}$ value for C-24-UB and C-24-UB-WO scenarios. The graphs show that to maintain $R_{event}$ < 1 on the urban buses when the infected subject is speaking, it is necessary to reduce the MRO to 40 persons and 23 persons for the scenarios with windows open and closed (corresponding to distances of approximately 0.5 m and 0.65 m), respectively. These values are approximately halved with respect to the MRO values considering $R_{rs}$ alone. Therefore, to avoid a reduction of the occupancy, it would be necessary to keep the windows open in urban buses and adopt other strategies such as wearing masks. Indeed, the occupancy imposed by the regulation (93 persons) would guarantee a $R_{event}$ < 1 when FFP2 masks are worn (both with windows closed or open; i.e. scenarios C-24-UB-FFP2 and C-24-UB-FFP2+WO) or when surgical masks are worn with the windows kept open (scenario C-24-UB-SM+WO). In contrast, with the windows closed, the MRO would be < 93 even if all the commuters were wearing surgical masks (scenario C-24-UB-SM with an MRO = 80). Once again, we highlight that the close proximity risk was assumed to be negligible when the infected person wears a mask; thus, for those scenarios, the $R_{event}$ (and the MRO) is only related to the room-scale risk.

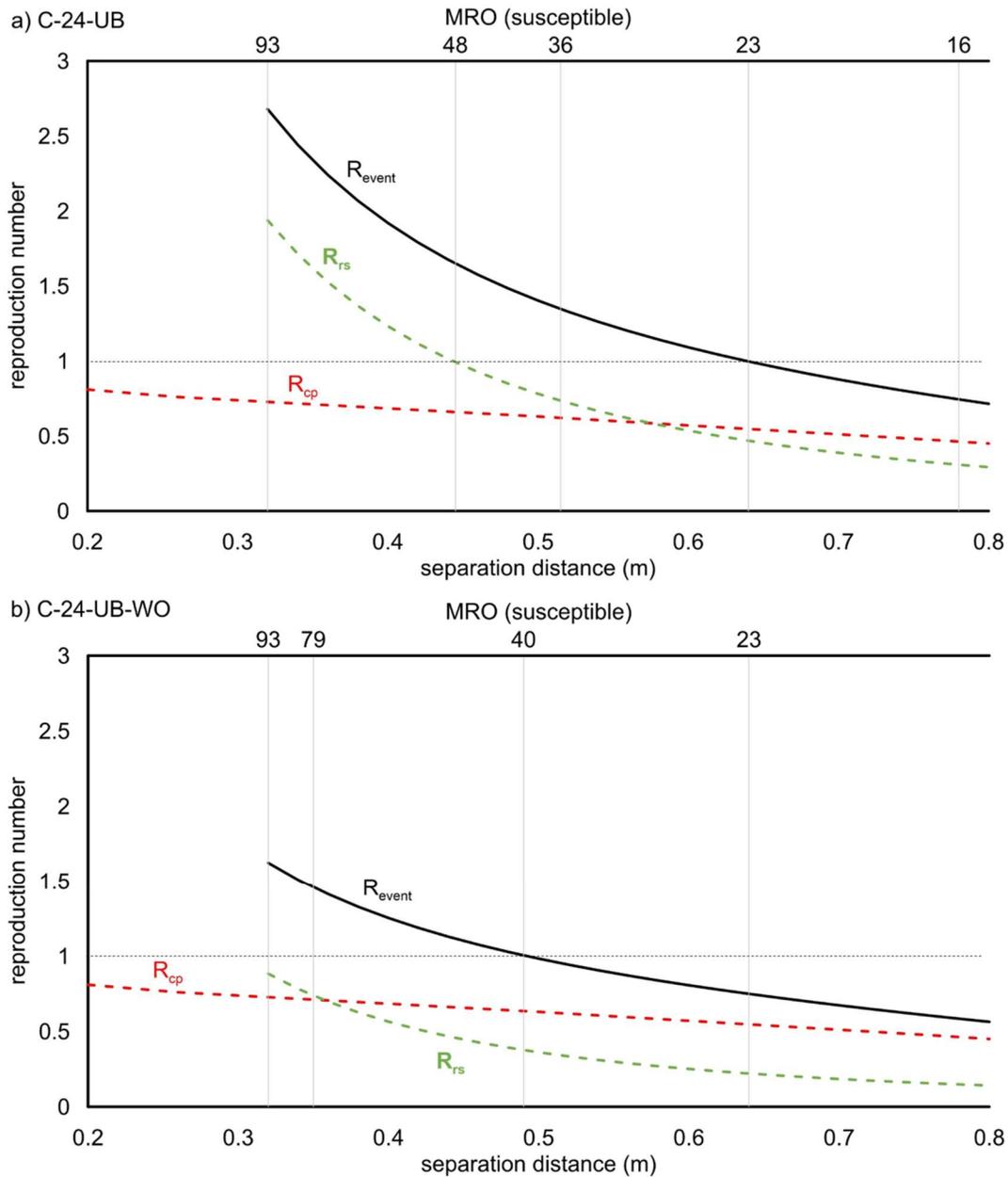

**Figure 2** - $R_{event}$ considering both close proximity ($R_{cp}$) and room-scale ($R_{rs}$) contributions for the C-24-UB (a) and C-24-UB-WO (b) scenarios. The equivalent maximum room occupancies (MROs) for maximum occupancy (93), $R_{rs} < 1$, and $R_{event} < 1$ are denoted by vertical lines.

### 3.2 Long-distance buses

For long-distance buses, the susceptibles travel for a long time (up to 480 min according to the investigated scenarios) in the same confined space with an infected subject, causing the $IR_{rs}$ to increase, and thus increasing the $R_{event}$. We point out that for long-distance buses, the $R_{event}$ is equal only to the room-scale reproduction number because the distances and the orientation between the commuters are such as to consider $R_{cp}$ negligible. Figure 3 shows an illustrative example of quanta concentration, $IR_{rs}$ and MRO trends for the scenario characterized by an infected commuter speaking for 60 min with no mitigation measures (case C-60-LDB). In this scenario, the $IR_{rs}$ reaches the maximum permitted value in 16 min (2%; i.e. 1 over 50 susceptible exposed persons), staying above that value for the entire travel time and thus not allowing full occupancy of the bus.

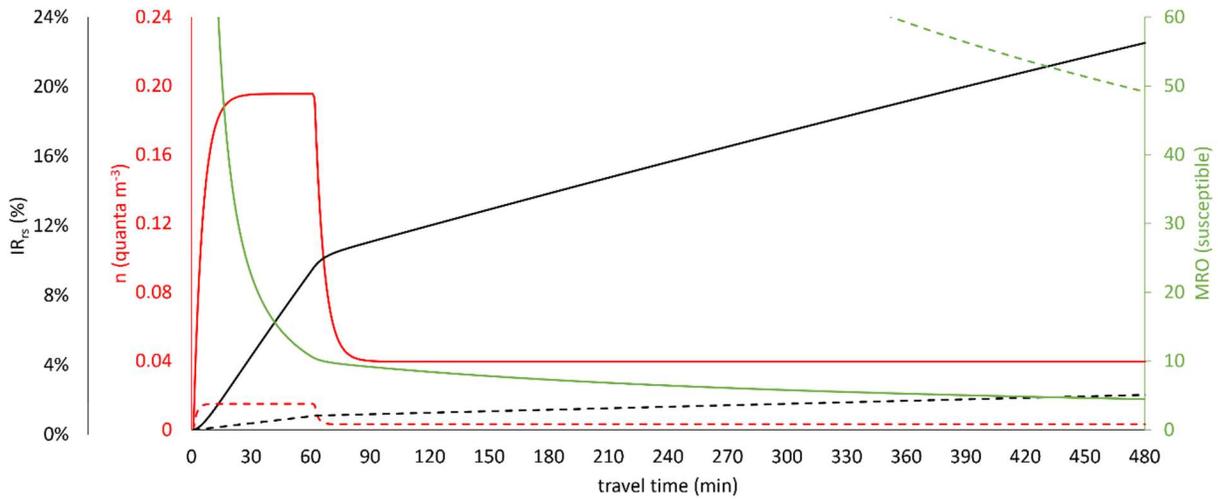

Figure 3 - Trends of quanta concentration (n), individual room-scale risk ($IR_{rs}$) and maximum room occupancy (MRO) for C-60-LDB (solid lines) and C-60-LDB-FFP2+M6 (dotted lines) scenarios.

Table 6 shows the required MRO to maintain $R_{event}$ < 1 for all the investigated scenarios for long-distance buses. For the C-60-LDB scenario, the MRO is extremely low even for a 1-h travel time (only 11 susceptibles could simultaneously share the bus) dropping to 4 commuters for an 8-h trip. The maximum occupancy of the bus is clearly related to the speaking time of the infected person during the trip (as also graphed in Figure 4). Indeed, when the speaking time is reduced to 30 min (C-30-LDB scenario) or 0 min (C-0-LDB, i.e. infected person only breathing) the MRO slightly increases. Nonetheless, full occupancy of the bus can be adopted for a 1-h travel time in the case of an infected individual only breathing. The use of surgical masks does not significantly improve the occupancy of the bus because, once again, full occupancy would be allowed only for a 1-h travel time as shown in Figure 4. Therefore, unless frequent breaks were taken during the trip to significantly lower the quanta concentration in the bus, further mitigation measures are needed to safely increase the number of commuters. In fact, the MRO data reported in Table 6 clearly highlight the fact that enhancing the ventilation through appropriate filtration of the recirculated air (M6 filter) and/or the use of more effective masks (FFP2) increases the MRO; indeed, when these two mitigation solutions are adopted simultaneously even for an infected subject speaking for 1 h, full occupancy would be permitted for very long trips. This is clearly illustrated in Figure 3 which shows the $IR_{rs}$ and MRO trends for the C-60-LDB-FFP2+M6 scenario: the individual risk $IR_{rs}$ remains below the maximum permitted value (2%) up to 457 min and the MRO for an 8-h travel time is 49 commuters (i.e. almost full occupancy).

Table 6 - Maximum room occupancy for long-distance buses to maintain a $R_{event}$ < 1 as a function of the scenarios investigated.

| Scenarios | Travel time | | | |
|---|---|---|---|---|
| | 60 min | 120 min | 240 min | 480 min |

| | | | | | |
|---|---|---|---|---|---|
| Base scenario | C-0-LDB | * | 25 | 13 | 7 |
| Commuter's speaking effect | C-30-LDB | 15 | 12 | 8 | 5 |
| | C-60-LDB | 11 | 8 | 6 | 4 |
| Filtration G3 & speaking effect | C-60-LDB-G3 | 12 | 10 | 7 | 5 |
| Surgical mask effect | C-0-LDB-SM | * | 42 | 21 | 11 |
| Filtration M6 effect | C-0-LDB-M6 | * | * | 30 | 15 |
| Filtration M6 & speaking effect | C-60-LDB-M6 | 26 | 21 | 15 | 10 |
| Surgical mask & speaking effect | C-60-LDB-SM | 18 | 14 | 10 | 7 |
| FFP2 & speaking effect | C-60-LDB-FFP2 | 51 | 40 | 30 | 20 |
| FFP2 & filtration M6 effect | C-60-LDB-FFP2+M6 | * | * | * | 49 |

\* The occupancy suggested by the ECE-R107 2015 regulation is satisfied.

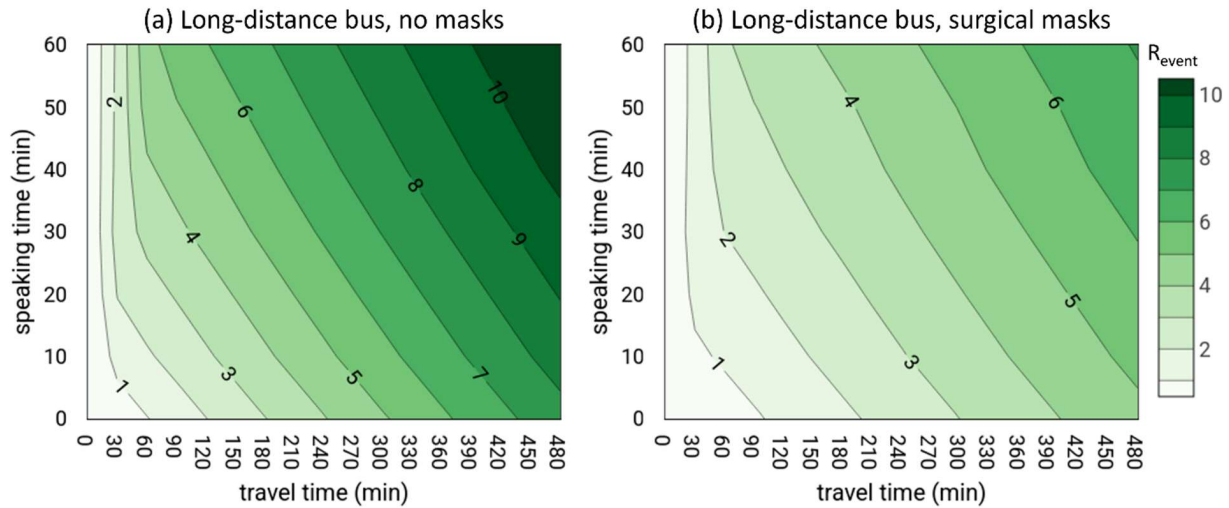

Figure 4 - $R_{event}$ as a function of speaking time (from 0 to 60 min) and travel time (from 0 to 480 min): (a) infected commuter speaking for the first minutes (0 to 60 min) and oral breathing for the rest of the time; (b) infected commuter speaking for the first minutes (0 to 60 min) and oral breathing for the rest of the time, with all commuters wear surgical masks.

When adequate filtration of the recirculated air and FFP2 masks are not adopted, the only solution to increase the MRO in long-distance buses is to reduce the number of susceptible people amongst those exposed; in other words, the fraction of the immune population (e.g. by vaccination) should be increased. To this end, Figure 5 shows the $R_{event}$ in long-distance buses as a function of the percentage of immunization and travel time (from 0 to 480 min) in the cases of breathing without mitigation measures (C-0-LDB), breathing with surgical masks (C-0-LDB-SM), speaking without mitigation measures (C-60-LDB), and speaking with surgical masks (C-60-LDB-SM). The figures indicate that, as expected, the presence of the immunes can reduce virus transmission; nonetheless, only a percentage of immunes higher than 90% would allow a $R_{event}$ < 1 in the case of 480-min trips with no masks. In fact, even if surgical masks were worn, a high percentage of immunes would still be required, i.e. at least 80% and 85% for oral-breathing and speaking activities, respectively.

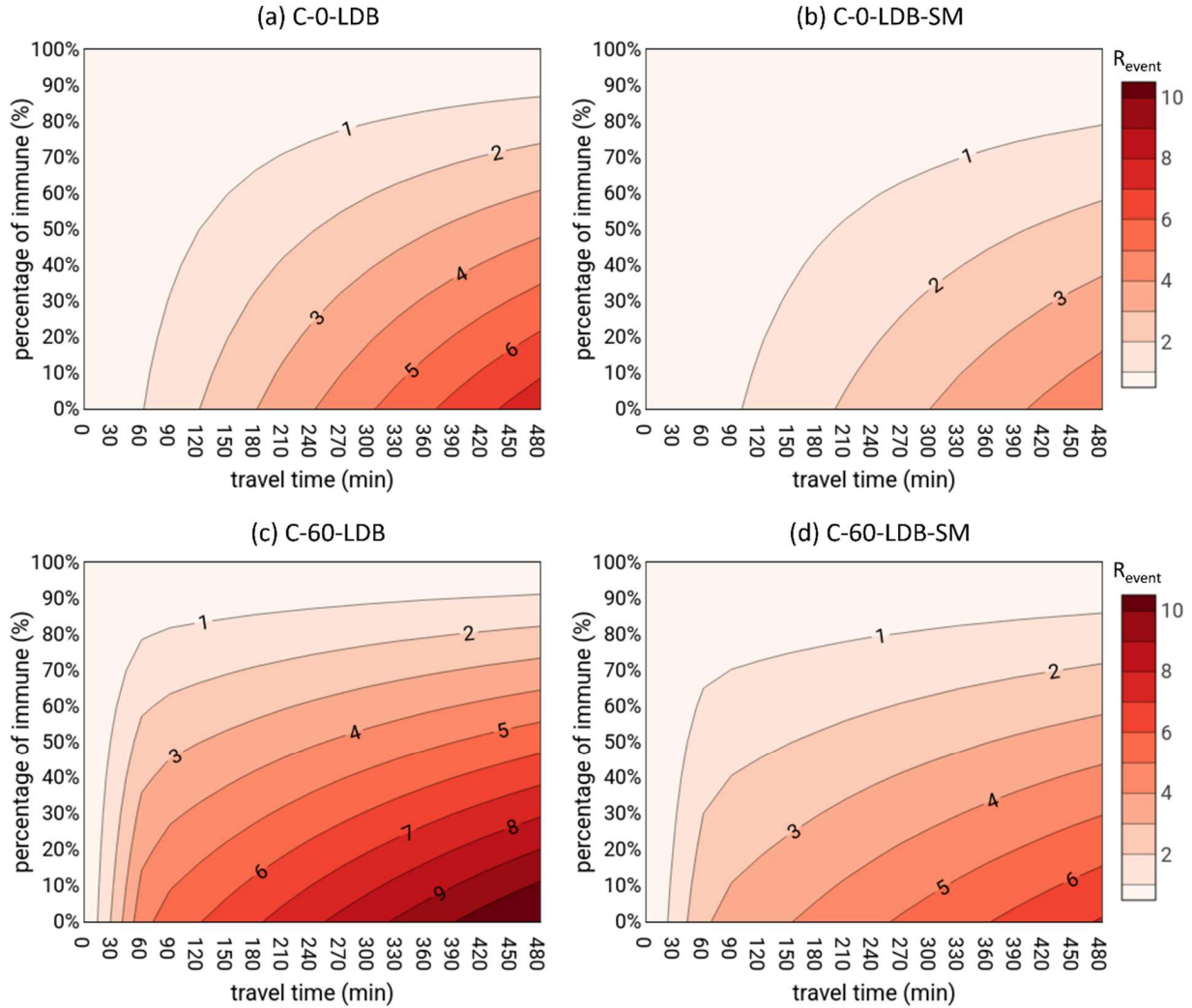

**Figure 5** - $R_{event}$ as a function of the percentage of immune individuals and travel time (from 0 to 480 min): (a) C-0-LDB, infected commuter standing for the whole trip oral breathing; (b) C-0-LDB-SM, infected commuter standing for the whole trip oral breathing and all commuters wear surgical masks; (c) C-60-LDB, infected commuter speaking for the first 60 min and oral breathing for the rest of the time and all commuters wear surgical masks; and (d) C-60-LDB-SM, infected commuter standing for the whole trip oral breathing and all commuters wear surgical masks.

## 4. Conclusions

This study evaluates the individual risk and the potential transmissibility (i.e. reproductive number, $R_{event}$) of SARS-CoV-2 infection in public buses, both in urban buses (characterized by a shorter exposure time but a higher crowding index) and long-distance buses (longer exposure time, lower crowding index). We considered the risk due to the proximity to the infected subject (close proximity contribution) and the risk related to the accumulation of virus-laden droplets in buses (also not in close proximity, i.e. room-scale contribution). Several typical scenarios in terms of ventilation, travel time, and expiratory activity of the infected subject are evaluated, as well as the adoption of mitigation strategies.

For urban buses, the contribution of close proximity to the individual risk is extremely high when the infected subject speaks for the entire travel time (up to 75% for full occupancy of the bus, i.e. at a separation distance of 0.32 m), thus significantly contributing to the reproductive number and, consequently, to the maximum occupancy of the bus in view of controlling the transmissibility of the pandemic. Indeed, the maximum occupancy to guarantee a $R_{event}$ < 1 (MRO) would be lower than the full occupancy of the bus both with the windows closed (measured ventilation rate of about 26 $h^{-1}$, MRO = 23 commuters) and with windows open (measured ventilation rate of about 65 $h^{-1}$, MRO = 40 commuters). To maintain a $R_{event}$ < 1 for full occupancy of the bus, masks should be adopted (FFP2 with the windows closed). For a breathing infected subject, the close proximity risk is negligible, and the room-scale contribution is 0.48%, thus guaranteeing a $R_{event}$ < 1 with full occupancy of the bus.

For long-distance buses, the close proximity contribution can be reasonably neglected due to the distances and orientation amongst the commuters; thus, the risk is only related to the room-scale contribution. The total exposure (travel) time and the adoption of mitigation solutions significantly affect the maximum occupancy of the bus. Reducing the speaking time and adopting frequent breaks during the trip represent very basic solutions that cannot always be applied. As an example, in the case of an infected person speaking for 1 h, only high quality filtration of the recirculated air and the simultaneous use of FFP2 masks would permit full occupancy of the bus up to almost 8 h; otherwise, an extremely high percentage of immunized persons (> 80%) would be required.